# Leakage Current Reduction Techniques in Poly-Si TFTs for Active Matrix Liquid Crystal Displays: A Comprehensive Study

Ali A. Orouji[1*], *Member*, *IEEE* and M. Jagadesh Kumar[2**], *Senior Member*, *IEEE*

[1]Department of Electrical Engineering, Semnan University, Semnan, IRAN

[2]Department of Electrical Engineering, Indian Institute of Technology, New Delhi 110016, INDIA

[*]Email: aliaorouji@ieee.org
[**]Corresponding author, Email: mamidala@ieee.org Fax: 91-11-2658 1264

*Abstract-* This paper critically examines the leakage current reduction techniques for improving the performance of poly-Si TFTs used in active matrix liquid crystal displays. This is a first comprehensive study in literature on this topic. The review assesses important proposals to circumvent the leakage current problem in poly-Si TFTs and a short evaluation of strengths and weaknesses specific to each method is presented. Also, a new device structure called the *Triple Gate* poly-Si TFT (TG-TFT) is discussed. The key idea in the operation of this device is to make the dominant conduction mechanism in the channel to be controlled by the accumulation charge density modulation by the gate (ACMG) and not by the gate-induced grain barrier lowering (GIGBL). Using two-dimensional and two-carrier device simulation, it is demonstrated that the TG-TFT exhibits a significantly diminished pseudo-subthreshold conduction leading to several orders of magnitude reduction in the OFF state leakage current when compared to a conventional poly-Si TFT. The reasons for the improved performance are explained.

I. INTRODUCTION

In recent years, polycrystalline silicon (poly-Si) thin film transistors (TFTs) have drawn much attention because of their wide application in active matrix liquid crystal displays (AMLCDs) [1], memory devices such as dynamic random access memories (DRAMs) [2], static random access memories (SRAMs) [3], [4], electrical programming read only memories (EPROMs) [5], electrical erasable programming read only memories (EEPROMs) [6], linear image sensors [7], thermal printer heads [8], photodetector amplifier [9], scanner, and neural networks [10]. Especially, the application of poly-Si TFTs in AMLCDs is the major reason to push the poly-Si TFTs technology programming rapidly. Due to the possibility of fabricating both NMOS and PMOS devices, thus enabling CMOS technology, and to the relatively large field-effect mobilities in both n- and p- channel devices compared to that of the amorphous silicon, poly-Si TFTs can be used to incorporate the integrated peripheral driving circuitry and switching transistor in the same substrate for AMLCDs [11]. This will reduce the assembly complication and cost dramatically. Also, the dimension of poly-Si TFTs can be made smaller compared to that of amorphous Si TFTs for high density, high resolution AMLCDs [12].

In polycrystalline silicon thin film transistors, grain boundaries and intragranular defects exert a profound influence on device characteristics and degrade carrier transport. Device characteristics such as threshold voltage, subthreshold swing, leakage current, mobility and transconductance will be poor compared with devices fabricated on single crystal silicon.

One of the important problems of poly-Si TFTs is large OFF state leakage current [13]. Several novel device structures have been reported in literature to circumvent the undesirable leakage current in poly-Si TFT devices. In spite of the importance associated with the leakage current reduction in poly-Si TFTs, there is no comprehensive review available to provide an understanding of the different leakage current reductions techniques. The aim of this work is to fulfill the above need. In this paper, in addition to reviewing important proposals, a new device structure called the triple-*gate* (TG) poly-Si TFT is discussed and its efficacy in suppressing leakage current in poly-Si TFTs is highlighted using two dimensional simulation.

## II. POLY-Si TFT FABRICATION METHODS

The conventional silicon technology based on crystalline silicon wafers is incompatible with the large-area electronics because of the high processing temperatures and the limited size of the wafers. This was the reason for the development of new techniques that could permit the silicon to be obtained by deposition over large surfaces at low temperature. On the other hand, the dangling bonds in the grain boundaries in the poly-Si film serve as the trapping centers. Free carriers (either electrons or holes) are contributed to the conduction band or valence band by substitutional dopant atoms located within the grains, just as the single crystal silicon. However, many of these free carriers are quickly trapped at low energy positions at the grain boundaries and consequently, cannot contribute to conduction [14], [15]. The trapped carrier will then deplete the charge nearby the grain boundaries. Because the traps located at the grain boundaries strongly influence the electrical properties of poly-Si and degrade the performance of the device, attempts have been made to modify or remove those grain boundaries. Traps are associated with the dangling bonds at the grain boundaries. Dangling bonds at a $Si$-$SiO_2$ interface are often passivated by terminating them with hydrogen atoms. Hydrogen can also passivate dangling bonds and other defects at grain boundaries [16]-[18], and reduce the number of active trapping states. As the number of trapped carriers decreases, the potential barrier associated with the grain boundary also decreases [2].

Therefore, crystallization of amorphous silicon (a-Si) is an excellent approach for making poly-Si films with large grain size and smooth surface. Various techniques have been used for crystallizing a-Si such as solid phase crystallization (SPC), laser induced crystallization (LIC) and metal induced crystallization (MIC). There are advantages and disadvantages for each technique.

The SPC technique is the lowest-cost and the most commonly used method. In this approach, a conventional thermal annealing in vacuum or in a neutral ambient (in argon or nitrogen gas) ensures the crystallization of the polysilicon films at an optimum temperature close to 600 °C suitable for fabrication of active-matrix on low cost glass substrates without mechanical deformation [19]. Most of the studies were focused on optimizing the deposition and crystallization conditions of polysilicon layers leading to larger grains. In other words, the crystallinity of polysilicon can be improved by adjusting the deposition pressure leading to an improvement of the electrical properties of the TFTs [20], [21]. The advantages of SPC are high reproducibility, good roughness control and uniformity [22]-[24]

Excimer-laser-annealing (ELA) is a promising laser induced crystallization (LIC) technique for TFT polysilicon layers [25]-[27], although it is a more expensive and a more difficult process to implement than the solid phase crystallization (SPC) process. In excimer-laser-annealing, a highly energetic radiation achieves the melting of the upper part of polysilicon layer. The crystallization occurs by displacement of the melting front to the surface. The maximum temperature at the back interface of the polysilicon layer remains low enough to avoid any deformation of the glass substrate. In spite of a high surface roughness [28], excimer laser annealed polysilicon layers usually offer a higher degree of crystallinity [29], [30] (a larger average grain size with a lower intra-grain defect density) than can be achieved using SPC polysilicon layers. The LIC method is not suitable for large area depositions due to the limited size of a conventional LIC beam. Therefore, a substrate typically requires several LIC scans; consequently, overlapping regions between different scans may cause device characteristics to vary depending on the scan patterns [29]. But, this technique can be used for low temperature processes.

When some metals are added into a-Si, the crystallization temperature can be lowered below 600 °C [31], [32] and this phenomenon is known as metal induced crystallization (MIC). However, the disadvantage of the MIC technique is that the crystallized Si films get contaminated with the metal. The solution to this problem is the metal induced lateral crystallization (MILC) technique. For palladium, MILC was shown to extend into a metal free, lateral area over one hundred microns, and the possibility of excluding metal contamination and obtaining large-grained poly-Si films was demonstrated [33]. Also, the first poly-Si TFTs formed and reported using MILC were made using Ni as the crystallization promoter [34]. The results exhibited good electrical characteristics for TFTs made using conventional SPC. Also, a reduction in the annealing time and temperature used for crystallization was achieved (i.e. 5 h at 500 °C for MILC when compared to 20 h at 600 °C in a typical SPC). Therefore, MILC offers a simple thermal process and is more compatible with glass substrates.

Recently, a hybrid process is proposed by Murley *et al*. in which a-Si is first converted to poly-Si using MILC method with Ni and then improved using excimer laser annealing [35]. This method is called Laser MILC or L-MILC.

It is worth noting that several groups [36]-[38] have reported that the use of polycrystalline silicon-germanium (poly-Si$_{1-x}$Ge$_x$) instead of poly-Si would significantly reduce the thermal budget of device fabrication. However, these completed devices did not exhibit acceptable performance as compared to the poly-Si devices thus preventing them from practical usage.

### III. LEAKAGE CURRENT REDUCTION TECHNIQUES

Various leakage current mechanisms in poly-Si TFTs have been reported in literature. Space-charge limited flow of holes from source to drain, thermal emission of carriers via grain boundary traps in the depletion region near the drain [39], [40], field enhanced thermal emission in the depletion region [39], parasitic bipolar effects [41], impact ionization in the drain depletion region, band-band tunneling in the depletion region, and field emission via grain boundary traps [13], [42], [43] are examples of these mechanisms. Therefore, the leakage current of poly-Si TFT will decrease if we can control the above mechanisms. It is worth noting that two factors play a significant role in increasing the OFF-state leakage current in poly-Si TFTs. First, the electric field due to the gate and drain voltages. When biasing the poly-Si TFTs in the OFF state with high source-drain voltages, high electric fields are formed near the drain and several field enhanced generation mechanisms control the leakage current. Second, the grain boundary trap density near the drain plays an important role. Poly-Si crystallization and hydrogen passivation are the two practical methods used to reduce the density of grain boundaries and therefore, reduce the leakage current [16]. The mostly commonly proposed structures are based on decreasing the electric field due to the gate and drain voltages. In this section, we review and assess important proposals to circumvent the leakage current in poly-Si TFTs and a short evaluation of strengths and weaknesses specific to each method is presented.

It is worth noting that the polarity between source and drain in a poly-Si TFT for AMLCD applications must be altered to reduce the DC stress of liquid crystal. Therefore, symmetrical poly-Si TFTs are required in AMLCD pixel control [44]. The asymmetric structures or misaligned processes may be responsible for the asymmetric electrical characteristics, such as threshold voltage shift and variation of the subthreshold slope.

### A. Lightly Doped Drain Structure

As mentioned above, an important approach for reducing leakage current is to decrease the electric field in the drain depletion region. This can be achieved using a lightly doped drain (LDD) structure as shown in Fig. 1 [45]-[47]. In this structure, $n^{-}$ regions are introduced between the channel and the $n^{+}$ source/drain regions. These LDD or offset regions can reduce the electric field near the drain. Some leakage current models for LDD poly-Si have been proposed [48]. These models show that the drain leakage current of a poly-Si TFT depends strongly on the drain voltage and temperature and weakly on gate bias. Therefore, the reduction rate in the leakage current is dependent on the doping and length of the offset regions. In a TFT structure implanted with $5 \times 10^{13}$ cm$^{-2}$ phosphorous ions in the 7 μm offset region length, the ON/OFF current ratio becomes more than one order of magnitude larger than that of a conventional TFT structure [45].

### B. Source overlap and a drain offset structure

It is clear that LDD structure suffers from ON current reduction due to the series resistance of the offset regions. To overcome this problem, the TFTs with a source overlap and drain offset [49]-[51] have been proposed. Fig. 2 shows a cross-sectional view of the bottom-gated polysilicon PMOS TFT with a source overlap and a drain off-set structure. In this structure, the OFF current is low because of the reduced electric field in the drain region due to the application of the drain offset region. However, the ON current is enhanced due to the presence of the source overlap region. Nonetheless, with increasing source overlap length, the channel length will decrease leading to a reduction in the channel punch-through breakdown voltage and gate breakdown voltage [49]. It is worth noting that the effect of parasitic capacitance from the source overlap should also be considered in the performance of this structure.

### C. Field-Induced-Drain (FID) Structure

Fig. 3 shows the fundamental structure of Field-Induced-Drain (FID) poly-Si TFT [52] in which the gate is divided to three sections: two sub gates and a main gate. With the application of a voltage to the sub gates, an inversion layer induced under the sub gates. This inversion layer decreases the electric field at the drain junction. It is necessary to make a low-resistive inversion layer by applying a high voltage to the sub gates. The high bias required for the sub gates is one of the drawbacks of this method. We can reduce the sub gate bias voltage using a thinner gate oxide or a higher dielectric constant material for the gate oxide. An ON/OFF current ratio of $10^7$ was successfully obtained with the FID TFT [52].

### D. N-P-N Gate Structure

In this method, as shown in Fig. 4, the gate of poly-Si TFT is composed of N-P-N junctions [53]. The main gate and the sub gates are composed of the $p^+$ poly region. Also, the gate voltage is applied to the main gate and the sub gate is connected to the source.

The key feature of this approach is that the vertical electric field between the drain and the gate reduces only the OFF state leakage current without significantly affecting the ON state current using the PN junction between source and gate. In the OFF state ($V_{Source}$ > $V_{Drain}$), the source voltage is applied to the $n^+$ gate through the forward biased PN junction. Therefore, the source voltage is isolated from the main gate due to the reverse biased PN junction between the $n^+$ gate and the $p^+$ main gate and so the source voltage prevents the increase of the vertical electric field between the drain and the gate. The leakage current of this structure is measured to be two orders of magnitude lower while the ON current is almost identical to the conventional device. An important advantage of this structure is that it does not need an additional mask step.

### E. Floating Sub-Gate Structure with Using Photoresist Reflow

Since the photoresist can act as a mask in ion implantation processes, some structures have been proposed for reducing the leakage current in poly-Si TFTs based on this idea [54], [55]. Fig. 5 shows a schematic feature of the proposed poly-Si TFT by Park *et al.* [54] with a self-aligned offset-gated structure by employing a photoresist reflow process. This device exploits the gate patterns to define the offset region, so that no additional mask step is required while eliminating any misalignment problem. The gate structure contains a main gate and two sub gates. The sub gates work as the extended part of the gate oxide and are fabricated by the photoresist reflow process [54]. Also, the gate bias is not applied to the sub gate so that the poly-Si channel under the offset oxide acts as an offset region. No requirement of an additional offset mask and good symmetrical electrical characteristics are the advantages of this structure.

### *F. Offset Gate Structure with Four Layers Gate*

Another offset gate structure is shown in Fig. 6 [56]. As can be seen from the figure, the four layers of the gate structure (gate poly-Si, gate oxide, upper poly-Si and buffer oxide) are used for making the offset regions. Compared with the conventional non-offset device, the leakage current deceases due to the reduction of the electric field in the drain depletion region. Some significant merits of this method are that the structure does not need an additional mask and a very thin film TFT device may be fabricated because the contact over-etch problem may be prevented.

### *G. Air Cavity Structure*

Applications of low dielectric constant materials such as air have been reported in VLSI technology [57]. Low dielectric constant materials are also used to reduce the interconnection capacitance and electric field. Based on the above idea, a poly-Si TFT structure with air cavities at the gate oxide edges has been proposed [58]. Fig. 7 (a) shows an air cavity TFT structure. Due to the low dielectric constant of air ($\varepsilon_r = 1$), the lateral and vertical electric fields are simultaneously decreased. It worth noting that an air cavity functions like a thick silicon dioxide ($\varepsilon_r = 3.9$) film with 3.9 times of the air cavity height as shown in Fig. 7(b). Therefore, the poly-Si region under the air cavity can be considered as an offset region and the gate edge over the air cavity serves as a field plate connected with the gate, so the proposed structure operates as a field induced drain (FID) poly-Si TFT discussed in section III-C.

### *H. High-k Spacer Offset-Gated Structure*

The schematic view of the self-aligned offset-gated poly-Si TFT with oxide, nitride or high-spacers [59] is shown in Fig . 8. This structure is proposed to reduce the OFF state leakage current and increase the ON-state current. In the off state, the lateral electric field is reduced due to the offset region between the drain and the gate. In the ON-state, due to the gate fringing field coupled through the high-$k$ spacer, a high vertical electric field is induced in the offset region underneath the spacer. Therefore, the electron density will increase leading to an inversion layer formation underneath the high-$k$ spacer. As a result, the series resistance of the offset region decreases. Table I shows the difference between the characteristics of different high-$k$ materials with the conventional poly-Si TFT. As can be seen from Table I, the leakage current for the devices with different spacer materials is almost same due to the similar offset length used. But, the ON/OFF ratio of the high-$k$ spacer is higher than the conventional (non-LDD) poly-Si TFT due to the induced inversion layer underneath the high-$k$ spacer. Also, the experimental results show that the ON state current at $HfO_2$ spacer offset-gated poly-Si TFT is about two times higher than that of the conventional oxide spacer TFT with the same leakage current [59]. The above device is more suitable for channel scaled-down system-on-panel applications.

### I. Vertical Bottom-Gate Structure

The channel length of vertical thin film transistors (VTFT) is determined by the thicknesses of $SiO_2$ or polysilicon films instead of the photolithographic limitations [60], [61]. Therefore, the VTFTs are suitable for high density integration. Fig. 9 shows a VTFT structure having an inherent off-set drain structure [62]. The device has two vertical channels with the source at the top and the offset drains at the two sides of the isolation oxide. The channel length of this device was determined by the thickness of bottom poly-Si gate and the length of the offset region was determined by the etching of the isolation oxide. Using a vertical structure, deep submicron channel lengths can be obtained even on large area wafers without having to face the photolithographic system limitations. Also, this structure does not need to any additional lithography step. The leakage current of the above TFT, however, should be reduced below 1 pA/μm using an optimized doping profile produced by a source/drain implantation for this technology to be viable.

### J. Gate Overlapped LDD (GOLDD) Structure

As mentioned before, the LDD or offset regions might degrade the device driving capability due to the large series resistance existing in these regions. In order to reduce the electric field at the drain without appreciable series resistance effects, gate overlapped lightly doped drain (GOLDD) structure has been proposed [63]-[65]. The GOLDD structure is shown in Fig. 10. As can be seen from the figure that n⁻ region is added to the structure. However, the GOLDD approach suffers inherent parasitic capacitance due to the large gate-LDD overlap.

### K. Amorphous Silicon Buffer Structure

The conventional LDD and offset poly-Si TFTs need additional mask steps to form the offset region. Also, the formation of the lightly doped n-region at the LDD structure complicates processing and increases the production cost. For solving these problems, a poly-Si TFT structure with a thin layer of hydrogenated amorphous silicon (a-Si:H) was proposed and is shown in Fig. 11 [66]-[68]. As can be seen from the figure, a very thin a-Si:H film (or buffer) was introduced to reduce the leakage current of the poly-Si TFT. This structure is fabricated with a $SiN_x$ gate insulator [69] and Ni-silicide source/drain contacts [70]. The effect of a-Si:H buffer can be explained using the band diagram shown in Fig. 12 for the poly-Si TFT with an a-Si:H buffer. The main idea for reduction of the leakage current in this approach is due to the greater band gap of the a-Si:H. The band gap of a-Si:H and poly-Si are 1.9 eV and 1.1 eV, respectively. Three current paths (A, B, and C) are shown in Fig. 12. The current along path C is due to the electron field emission for the $SiN_x$/poly-Si TFT that can be suppressed by the presence of a thin and large band gap of the a-Si:H buffer. A and B paths show the thermionic field emission in the a-Si:H and poly-Si region. It is worth noting that the leakage current of an a-Si:H TFT [71] is much smaller than that of a poly-Si. In other words, the current 'A' is much smaller than current 'C'. Therefore, the leakage current of the poly-Si TFT with a very thin a-Si:H buffer is smaller than that of the conventional poly-Si TFT.

## IV. MODIFIED CHANNEL CONDUCTION MECHANISM

In active matrix liquid crystal display (AMLCD) applications, circuit speed is often limited by the relatively low effective carrier mobility in the polysilicon channel. However, for high speed applications, short channel poly-Si TFTs are desired since the propagation delay is approximately proportional to the square of the channel length.

The transfer characteristic of the conventional SOI MOSFET has a steep subthreshold slope and dominant conduction mechanism is due to the inversion charge density modulated by gate (ICMG) [72]. On the other hand, there are two conduction regions in the transfer characteristic of a poly-Si TFT, as shown in Fig. 13, known as subthreshold region and pseudo-subthreshold region and the transition from the pseudo-subthreshold to the turn-on region is much more gradual [73], [74]. The dominant conduction mechanism of a poly-Si TFT is due to the gate-induced grain barrier lowering (GIGBL). To realize a diminished pseudo-subthreshold conduction in the operation of a polysilicon TFT, a new structure known as *Triple Gate* Poly-Si TFT (TG-TFT) has been proposed by us [75].

The key idea behind this approach is to modify the channel potential so that the channel conduction is controlled by the accumulation (like inversion in SOI-MOSFET) charge density modulation by the gate (ACMG) and not by GIGBL. By converting the dominant conduction mechanism in a poly-Si TFT from GIGBL to ACMG, due to steep subthreshold slope in the ACMG mechanism, the leakage current of the device is expected to reduce. As can be seen from Fig. 14, the front gate of TG-TFT structure consists of two p$^+$-poly side gates and the n$^+$-poly main gate. We have considered one grain boundary in the channel of the poly-Si TFT [76]-[78] for proving the validity of our approach. It is worth noting that it is possible to create devices where only a single or small number of discrete grain boundaries exist in the channel of the poly-Si TFT using modern metal-induced lateral crystallization (MILC) or excimer laser annealed (ELA) methods to control the grain growth [76], [77]. The simulation (structure) parameters are given in Table II and the side and main gate lengths are considered equal.

A typical MEDICI [79] simulated 2-D conduction band potential distributions for the conventional poly-Si TFT (C-TFT) for the drain to source voltage $V_{DS}$ = 10 mV and 1 V are shown in Fig. 15. It can be seen that a potential barrier (central barrier) is formed at the GB because the carriers are immobilized by the traps due to the strain and the dangling bonds located at the grain boundary [14], [15]. Therefore, the dominant conduction mechanism of the C-TFT is determined by GIGBL. But, in the proposed TG-TFT due to its triple-gate structure, in addition to the central barrier, two extra barriers (side barriers) are created in the side gate regions due to the work-function difference between the side gate and the main gate as shown in Fig. 16 for (a) $V_{GS}$ = 0 V and $V_{DS}$ = 10 mV and (b) $V_{GS}$ = 0 V and $V_{DS}$ = 1 V. The side and central barriers not only differ in their height but also in their shape. Since the central barrier does not play any significant role in the presence of the side barriers, the dominant conduction mechanism of the TG-TFT should now be controlled by the side barriers. In that case, we should have a steep subthreshold slope in the transfer characteristic of the device just as is observed in a typical single crystal SOI MOSFET. With increasing gate voltage, the height of the side barrier will decrease and at some critical gate voltage ($V_{CGS}$), the side barrier height will become equal to the central barrier height as shown in Fig. 17 for (a) $V_{DS}$ = 10 mV and (b) $V_{DS}$ = 1 V. The critical gate voltage $V_{CGS}$ increases from 0.52 V (Fig. 17(a)) to 0.75 V (Fig. 17(b)) when the drain voltage is increased from 10 mV to 1 V. After this critical gate voltage ($V_{CGS}$) condition is reached, the channel conduction mechanism will be determined by GIGBL.

In Fig. 18, the transfer characteristics of the TG-TFT are compared with that of the C-TFT and the single crystal SOI MOSFET for (a) $V_{DS}$ = 10 mV and (b) $V_{DS}$ = 1 V. We notice from this figure that as speculated above, for all gate voltages less than the critical gate voltage $V_{CGS}$ (= 0.52 V when $V_{DS}$ = 10 mV and 0.75 V when $V_{DS}$ = 1V ), the subthreshold slope of the TG-TFT is very steep similar to that commonly observed in SOI MOSFETs. For gate voltages greater than $V_{CGS}$, the transfer characteristic of the TG-TFT matches with that of the C-TFT since now the height of the central barrier is larger than that of the side barriers. Therefore, it is clear that because of the steep subtrheshold slope, the TG-TFT will have several orders of magnitude lesser off-state leakage current when compared to the C-TFT. This has become possible by nullifying the effect of the central barrier associated with the grain boundary on the channel conduction mechanism so that the pseudo-subthreshold region is almost eliminated.

The value of critical gate voltage $V_{CGS}$ is very important in controlling the pseudo-subthreshold region and hence the reduction in the off-state leakage current. An important parameter that determines the value of $V_{CGS}$ is the work function of the side gate ($\varphi_{Msg}$). Fig. 19 shows the transfer characteristic of the TG-TFT for different work functions of the side gate region. It can be seen from the figure that as the work function of the side gate decreases, the critical gate voltage $V_{CGS}$ will reduce forcing the behavior of the TG-TFT approach that of the C-TFT. In addition to the shift in the critical voltage, the ON current of the TG-TFT will also reduce. This is because if the work function of the side gate decreases for a given work function of the main gate, the height of the side barriers will also decrease. Therefore, it is very important to choose appropriate work function for the side gate for a given main gate work function ($\varphi_{Mmg}$).

## V. CONCLUSIONS

In sections 2 and 3 of this paper, the fabrication methods and leakage current reduction techniques in poly-Si TFTs have been analyzed and recent attempts to alleviate the leakage current problems have been reviewed. Specific strengths and weaknesses of the different approaches have been discussed. LDD or Offset structure is a popular way but that degrades the device driving capability due to the large series resistance existing in the offset regions. The GOLDD structure can solve the series resistance problem of the offset structure but that suffers inherent parasitic capacitance due to the large gate-LDD overlap. *Triple Gate* (TG) poly-Si TFTs promise a significant suppression of the leakage current in the channel. The efficacy of the TG structure in suppressing leakage current is assessed using two-dimensional simulation. Simulation studies demonstrate a diminished pseudo- subthreshold region and a huge leakage current reduction in a TG-TFT structure.

It is worth noting that the gate of the TG-TFT structure can be fabricated using the same experimental procedure as reported for Electrically Shallow Junction MOSFET (EJ-MOSFET) in [81]-[83] and this concept of modified channel conduction mechanism can be used even for asymmetrical double gate polysilicon TFTs [84].

**Table Captions**

Table I        Comparison between the device characteristics of different poly-Si TFT structures with W/L = 5μm/2μm, $V_g$(ON) = 20 V, $V_g$(OFF) = -10 V. The permittivity of $HfO_2$, nitride and oxide are 20, 8, 3.9, respectively [59].

Table II      Parameters for TG-TFT structure used for MEDICI simulation

**Figure Captions**

Figure 1      Cross section of an LDD or offset structure poly-Si TFT [45].

Figure 2      Cross sectional view of poly-Si TFT with a source overlap and a drain offset [49].

Figure 3      Fundamental Field-Induced-Drain (FID) poly-Si TFT structure [52].

Figure 4      Schematic view of the P-N-P gate structure [53].

Figure 5      Schematic view of the self-aligned offset gate with a floating sub-gate [54].

Figure 6      Cross-section of the offset gate structure with a buffer oxide [55].

Figure 7      The poly-Si TFT with air cavity, (a) schematic view, (b) equivalent structure [58].

Figure 8      Cross-sectional view of the high-*k* spacer poly-Si TFT structure [59].

Figure 9      Schematic view of the vertical bottom-gate poly-Si TFT [62]

Figure 10     Schematic of the gate overlapped lightly doped drain (GOLDD) structure [63].

Figure 11     Cross-sectional view of a poly-Si TFT with an a-Si:H buffer [68].

Figure 12     Band diagram in the poly-Si TFT with an a-Si:H buffer [68].

Figure 13     Pseudo-subthreshold conduction in a conventional poly-Si TFT [74].

Figure 14     Cross-sectional view of the TG-TFT [75].

Figure 15     Conduction band potential distribution for a C-TFT (a) with $V_{GS}$ = 0 V and $V_{DS}$ = 10 mV, (b) with $V_{GS}$ = 0 V and $V_{DS}$ = 1 V.

Figure 16     Conduction band potential distribution for a TG-TFT (a) with $V_{GS}$ = 0 V and $V_{DS}$ = 10 mV, (b) with $V_{GS}$ = 0 V and $V_{DS}$ = 1 V.

Figure 17     Conduction band potential distribution for a TG-TFT (a) with $V_{GS}$ = 0.52

V and $V_{DS}$ = 10 mV, (b) with $V_{GS}$ = 0.75 V and $V_{DS}$ = 1 V.

Figure 18    A comparison of the transfer characteristics of TG-TFT, C-TFT, and C-SOI structures for (a) $V_{DS}$ = 10 mV and (b) $V_{DS}$ = 1 V.

Figure 19    Transfer characteristics of the TG-TFT structure for different work functions of the side gates.

Table I

| Device Type | $\mu_{FE}$ (cm²/V.S) | $I_{OFF}$ (pA/μm) | $I_{ON}/I_{OFF}$ ×10⁶ |
|---|---|---|---|
| Non-LDD | 63 | 36.2 | 1.15 |
| Oxide Spacer | 65 | 16.4 | 1.58 |
| Nitride Spacer | 69 | 16.4 | 1.82 |
| HfO₂ Spacer | 72 | 16.4 | 2.51 |

Table II

| | |
|---|---|
| Doping in n⁺ source/drain | $1\times10^{19}$ cm⁻³ |
| Trapping density at the grain boundary states | $1\times10^{13}$ cm⁻² |
| Electron trap energy relative to the conduction (band) | 0.51 eV |
| Hole trap energy relative to the (valence band) conduction | 0.51 eV |
| Capture rate for electrons and holes | $1\times10^{-8}$ cm³/sec [80] |
| Width of the grain boundary | 10 nm |
| Silicon thin film thickness | 50 nm |
| Gate oxide thickness | 10 nm |
| Channel length L | 0.4 μm |
| P⁺-poly gate work function | 5.25 eV |
| N⁺-poly gate work function | 4.17 eV |

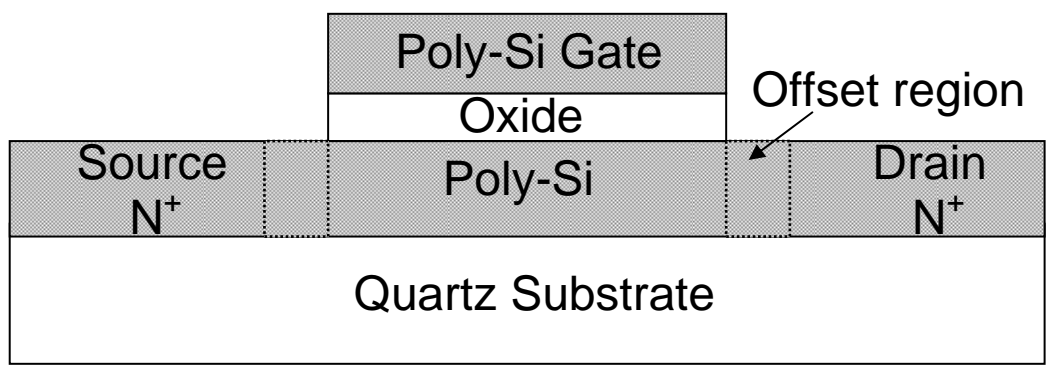

Fig. 1.

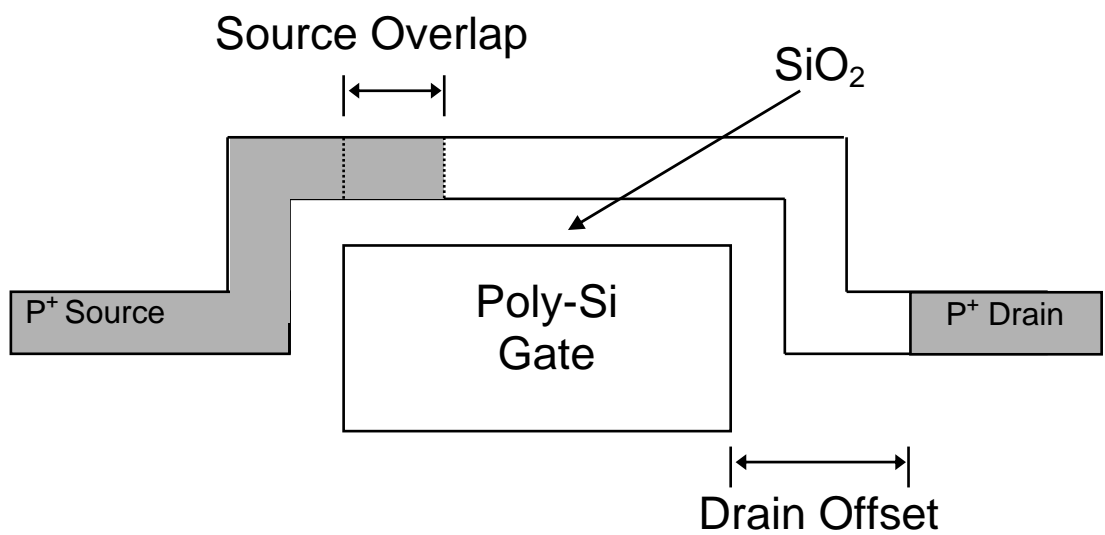

Fig. 2.

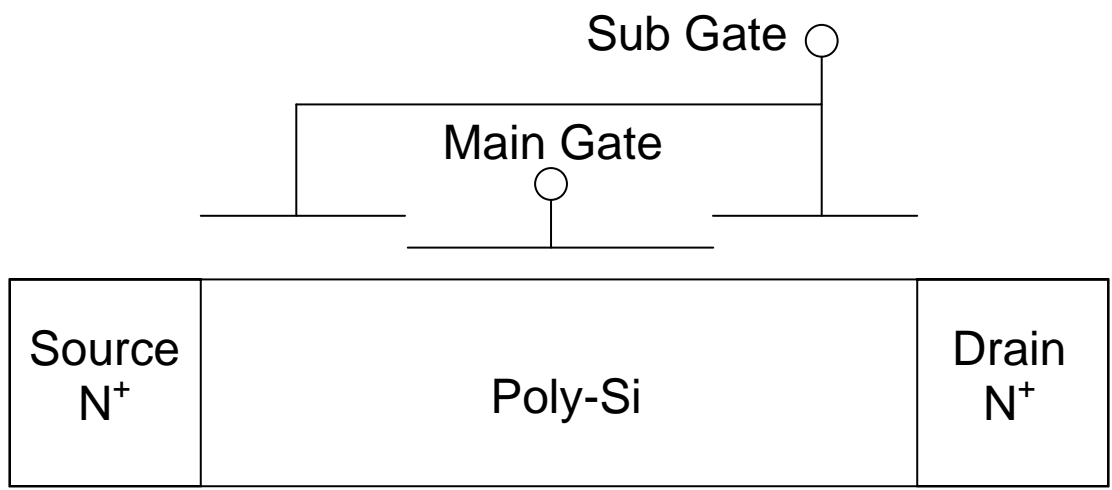

Fig. 3.

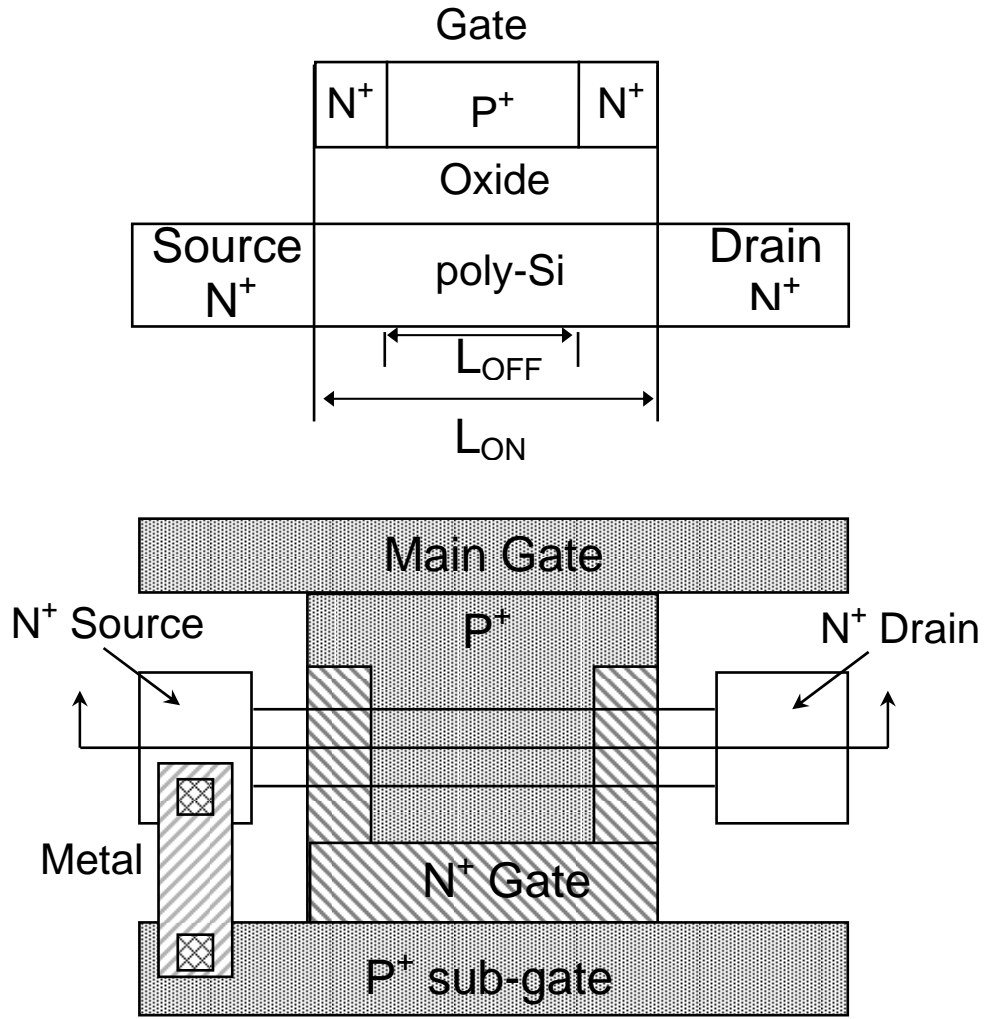

Fig. 4.

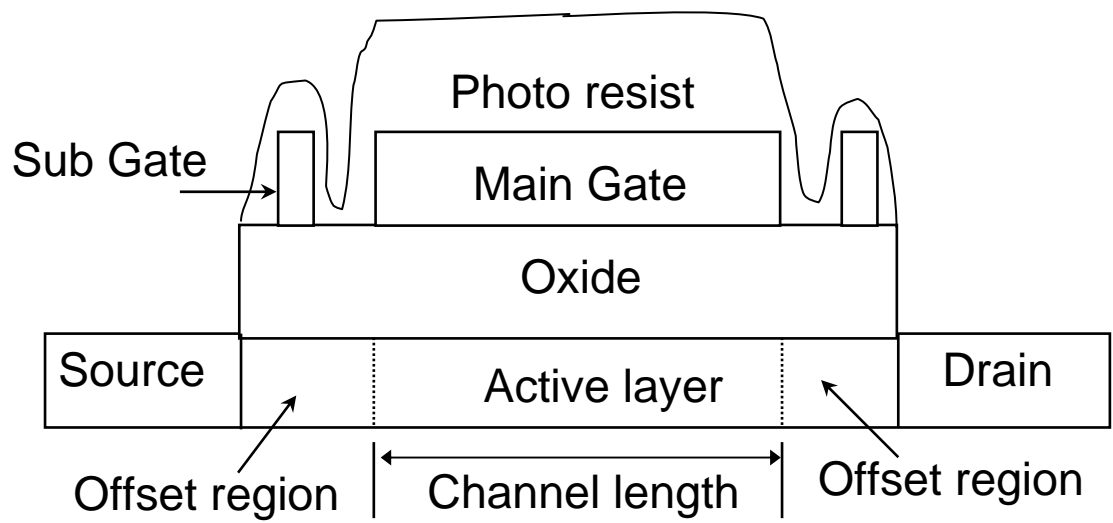

Fig. 5.

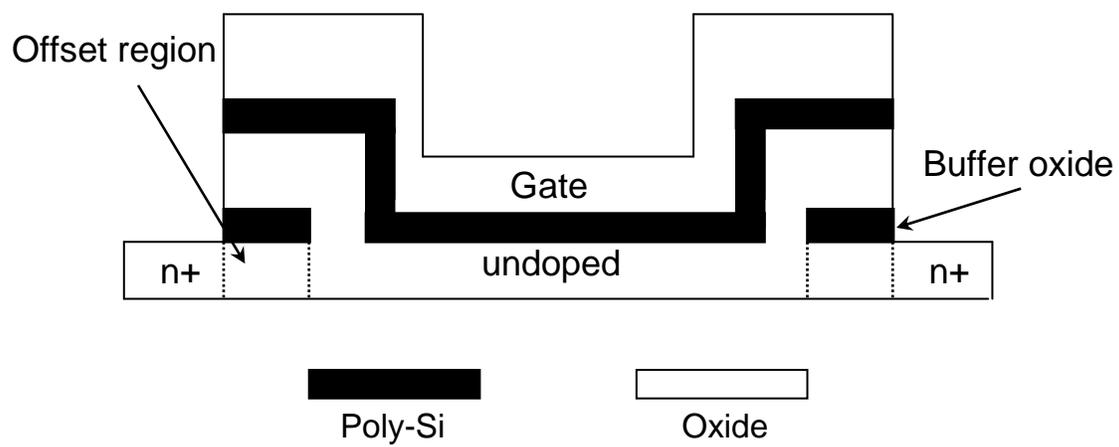

Fig. 6.

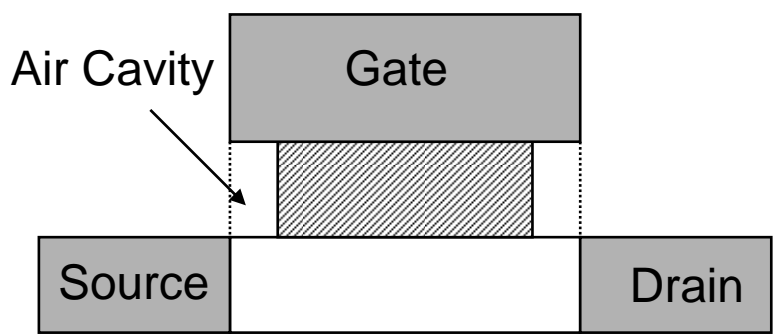

Fig. 7 (a)

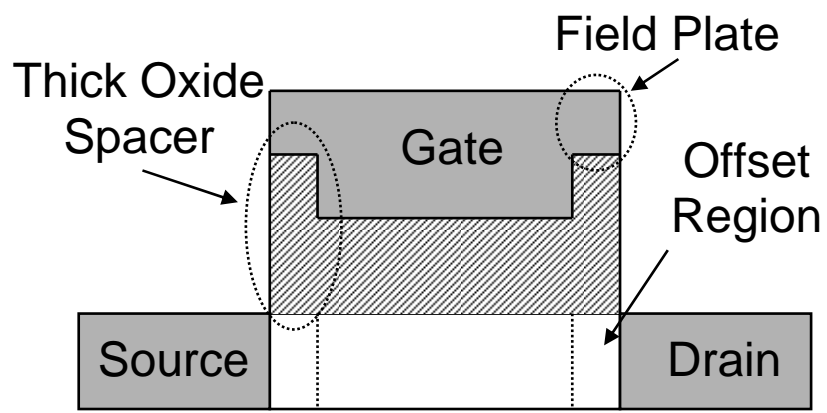

Fig. 7 (b)

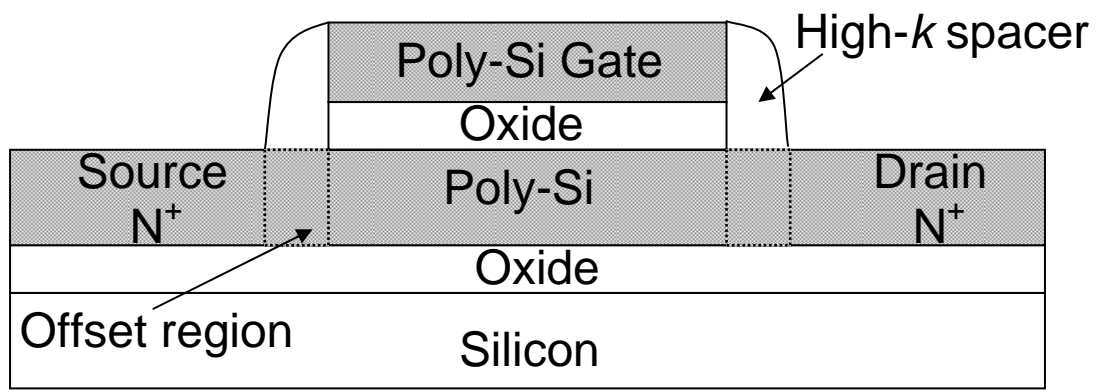

Fig. 8.

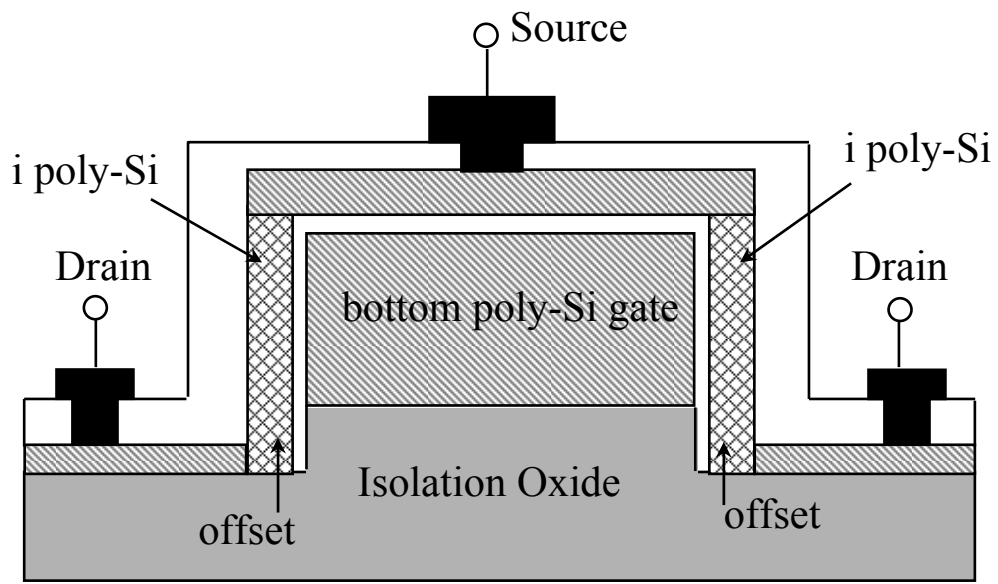

Fig. 9.

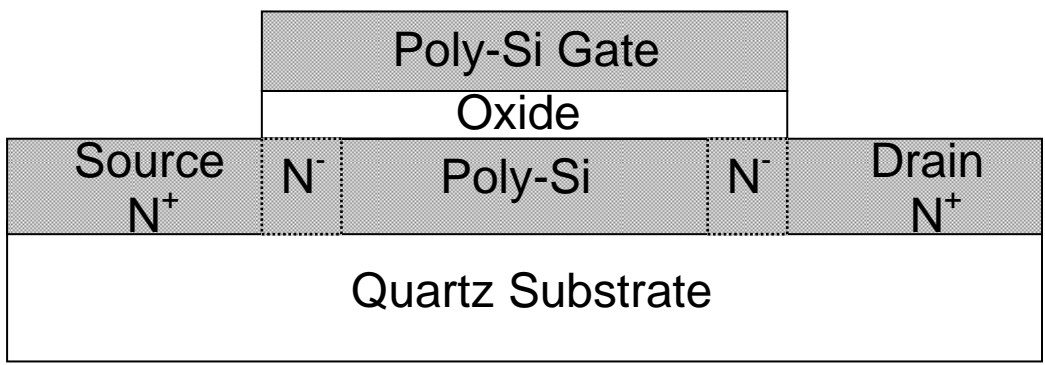

Fig. 10.

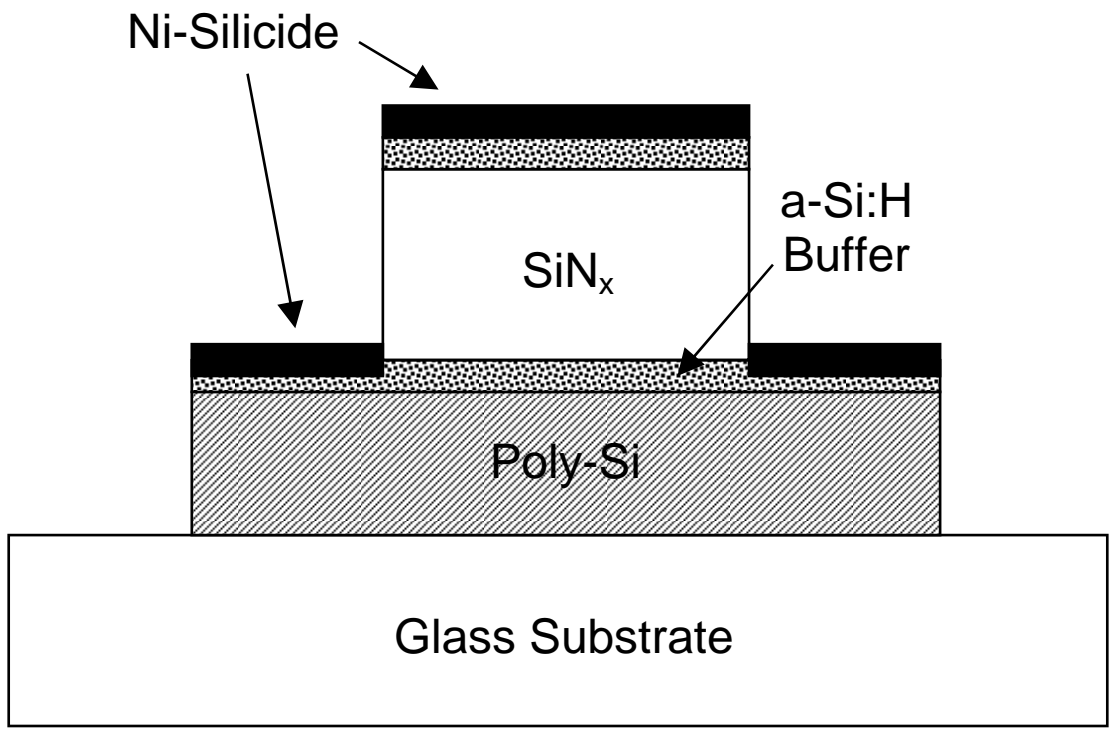

Ni-Silicide

a-Si:H
Buffer

SiN_x

Poly-Si

Glass Substrate

Fig. 11

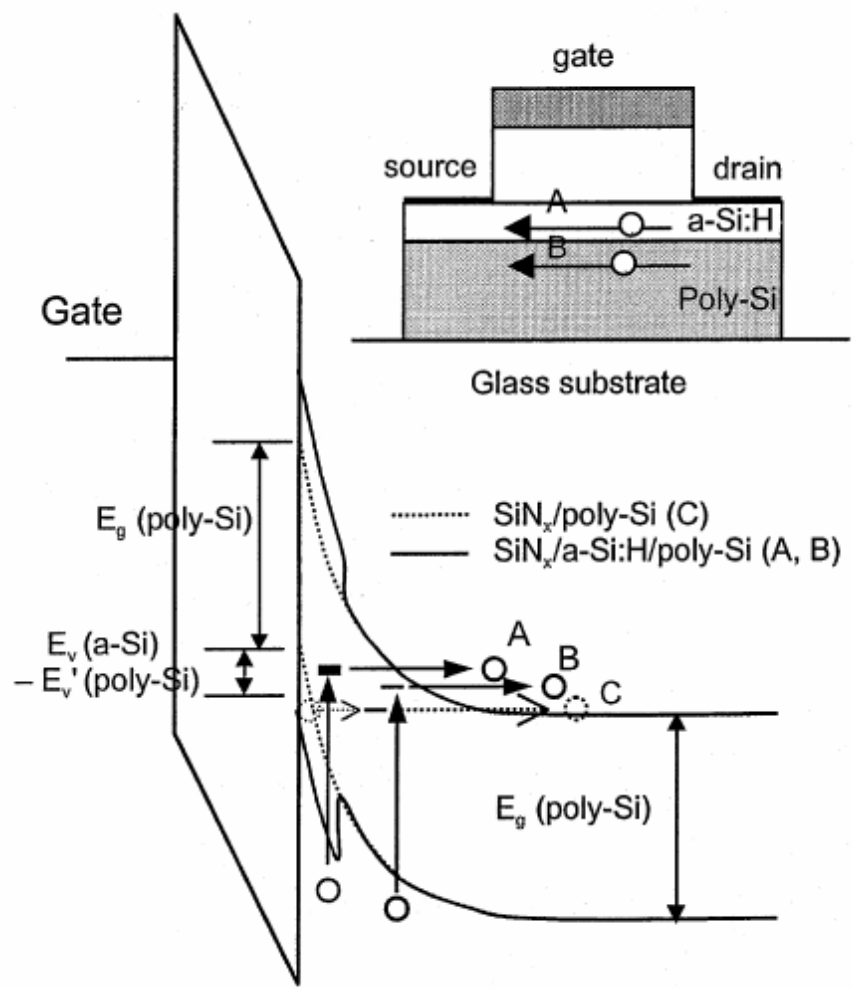

Fig. 12

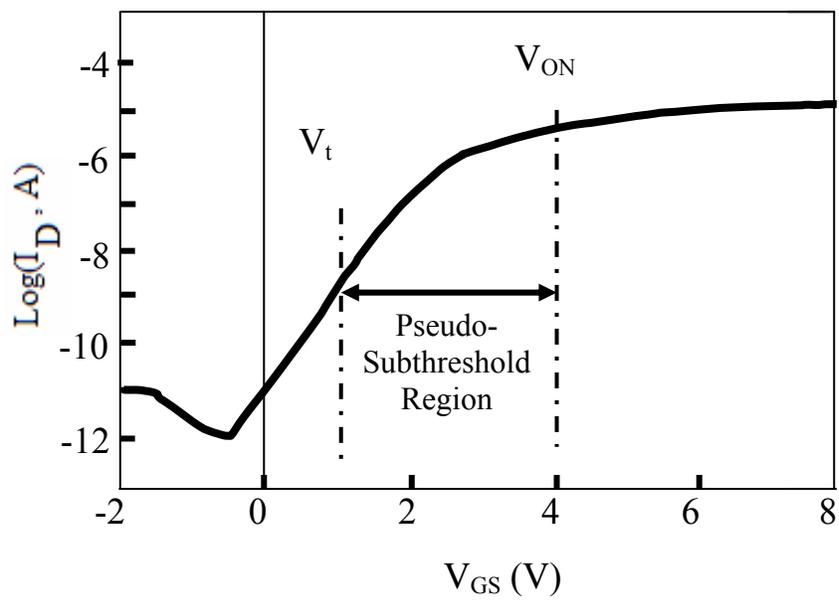

Fig. 13

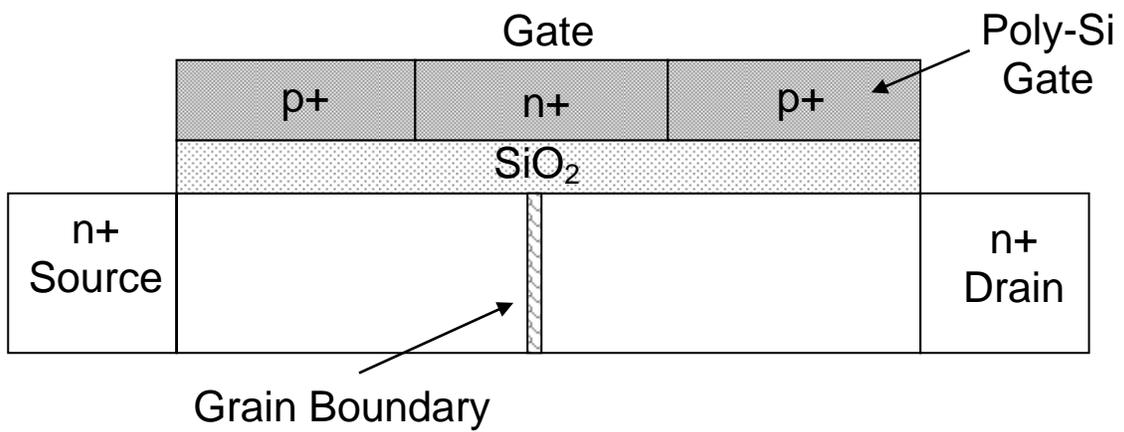

Fig. 14.

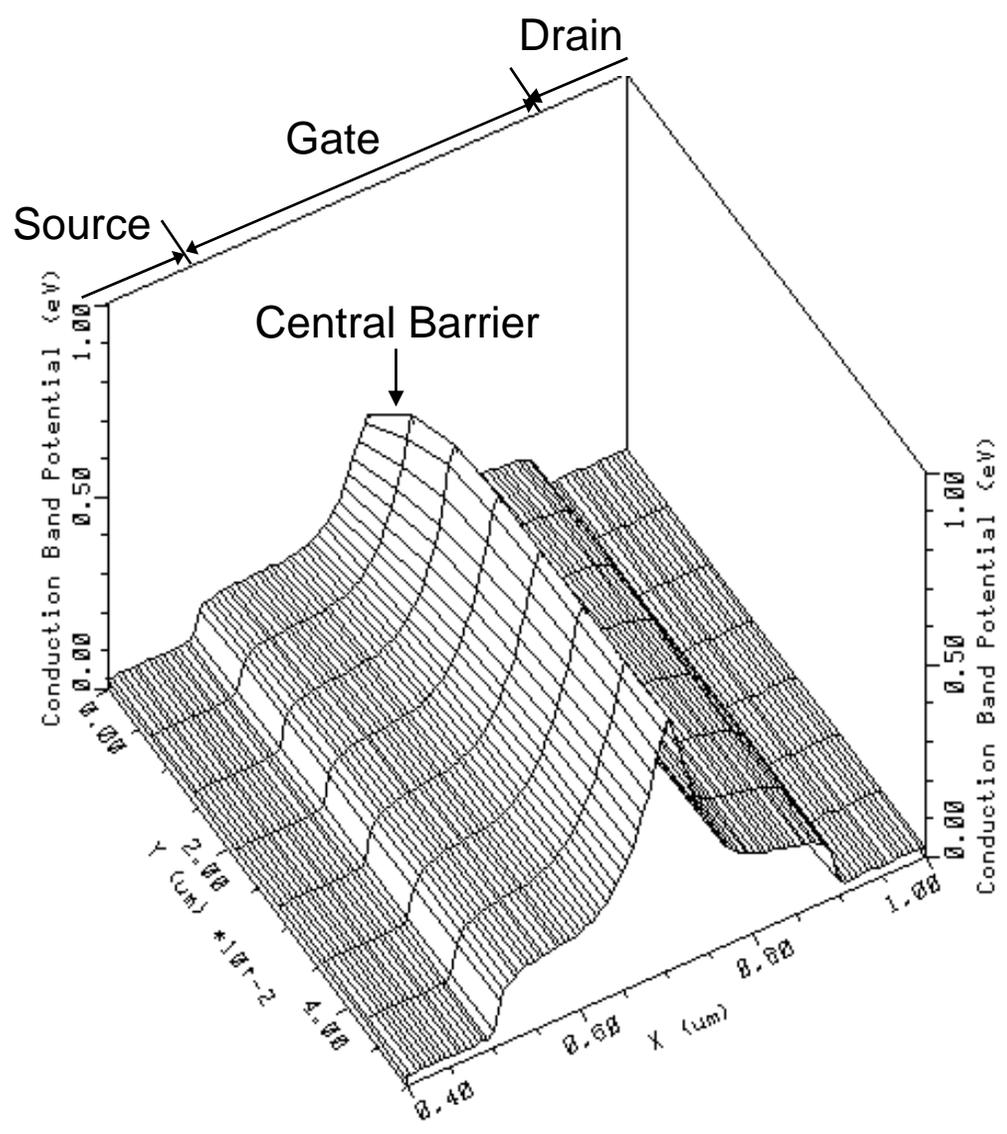

Fig. 15 (a)

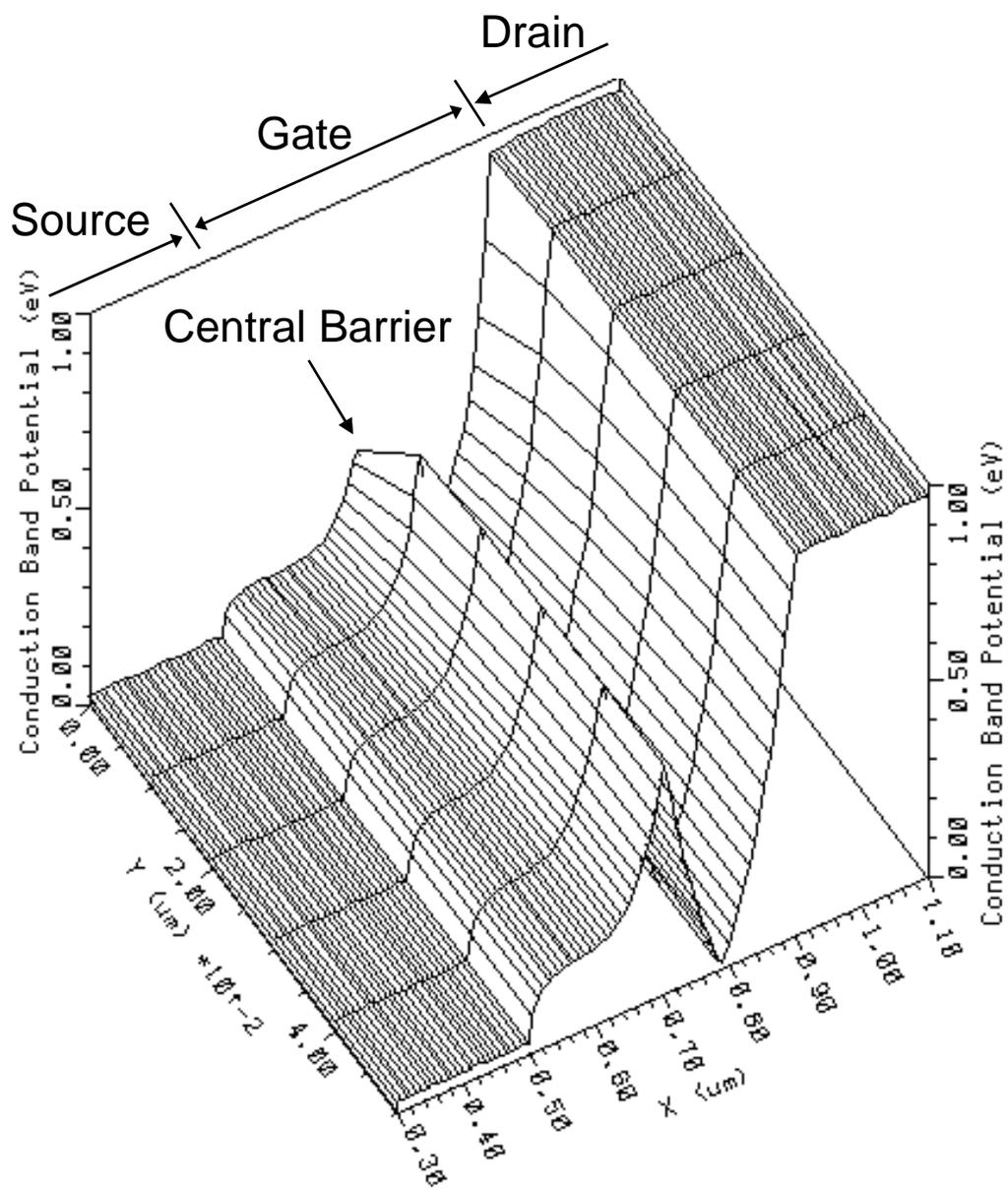

Fig. 15(b)

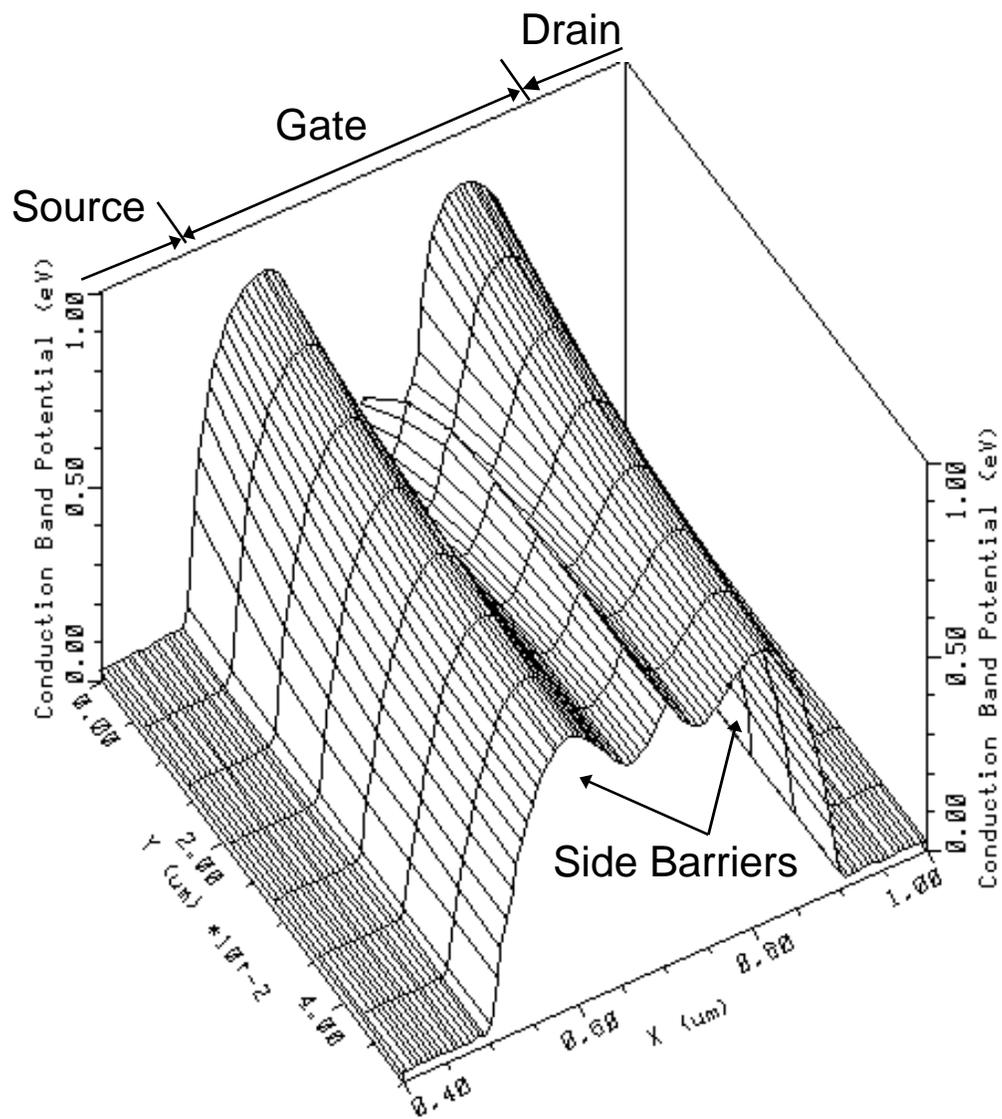

Fig. 16 (a)

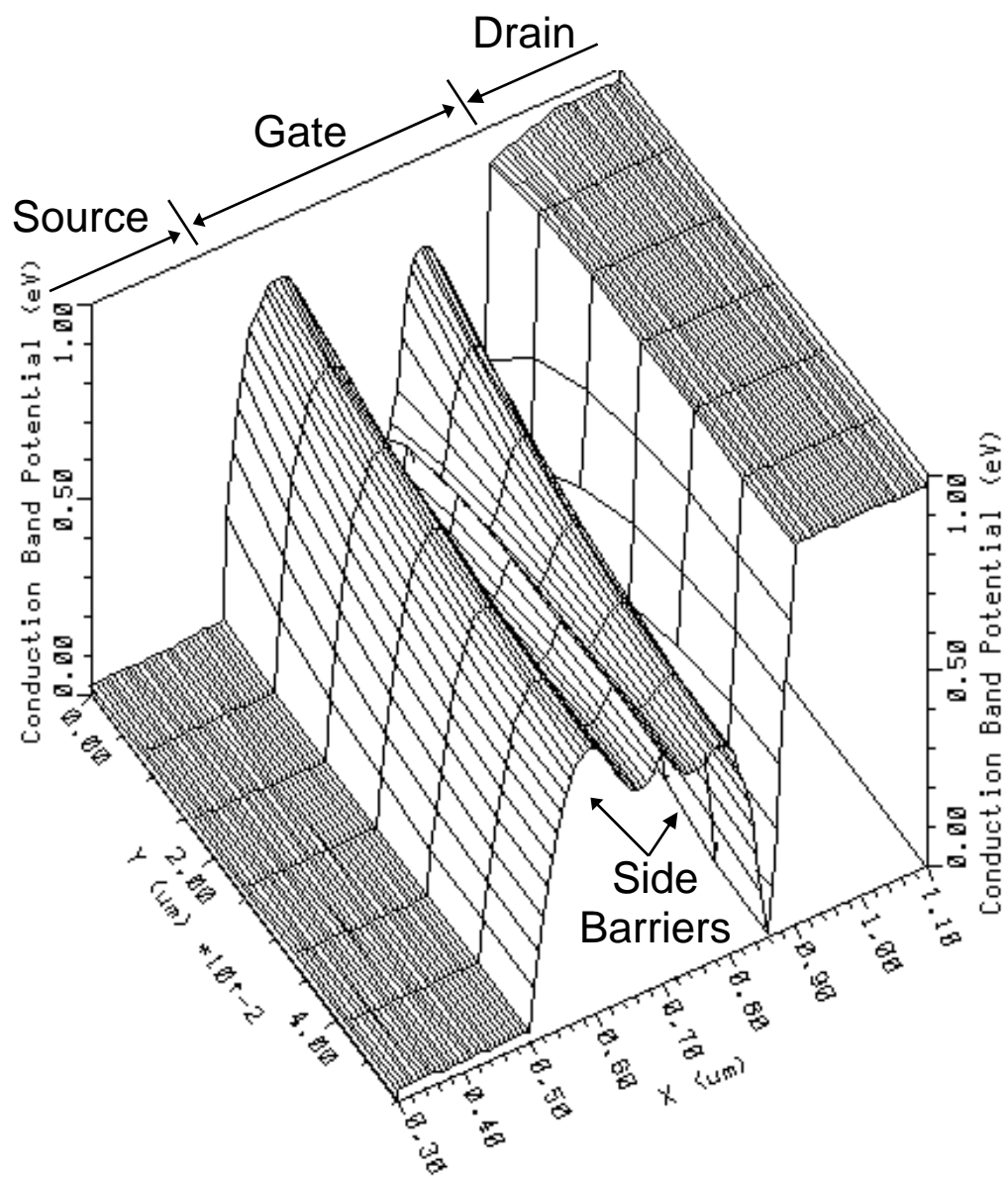

Fig. 16 (b)

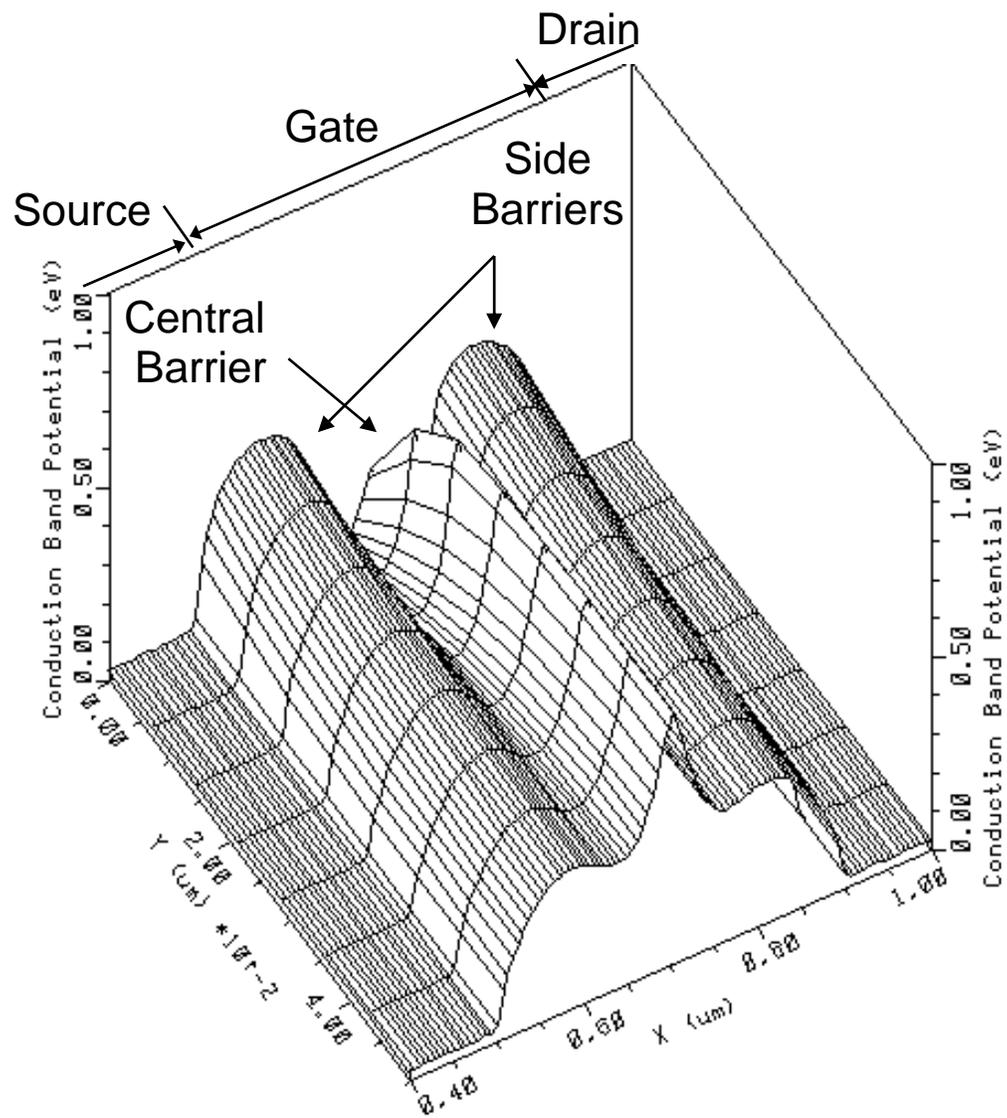

Fig. 17(a)

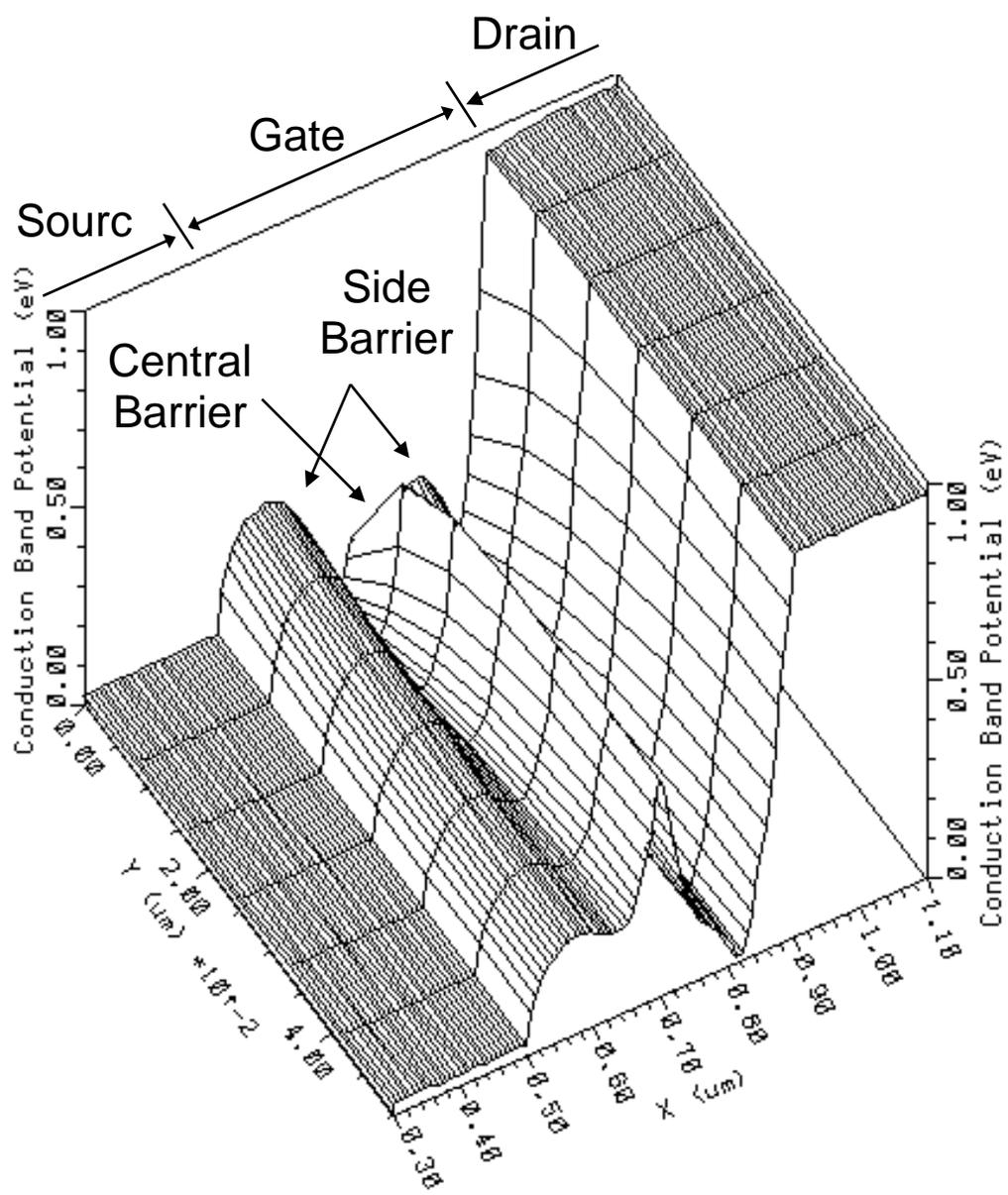

Fig. 17(b)

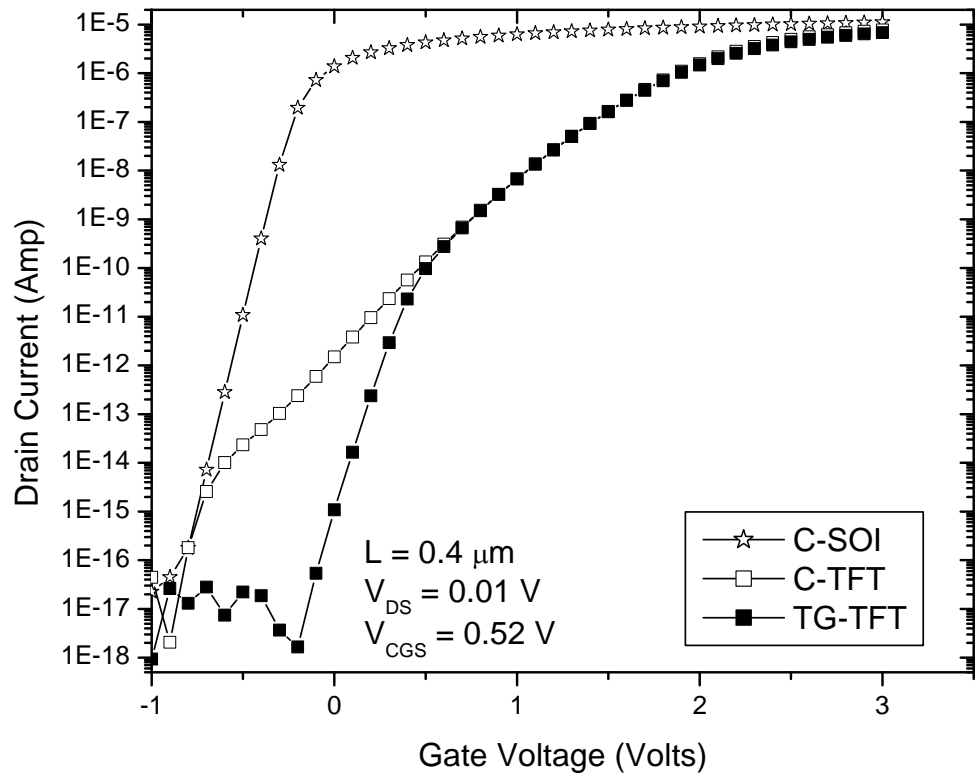

Fig. 18 (a)

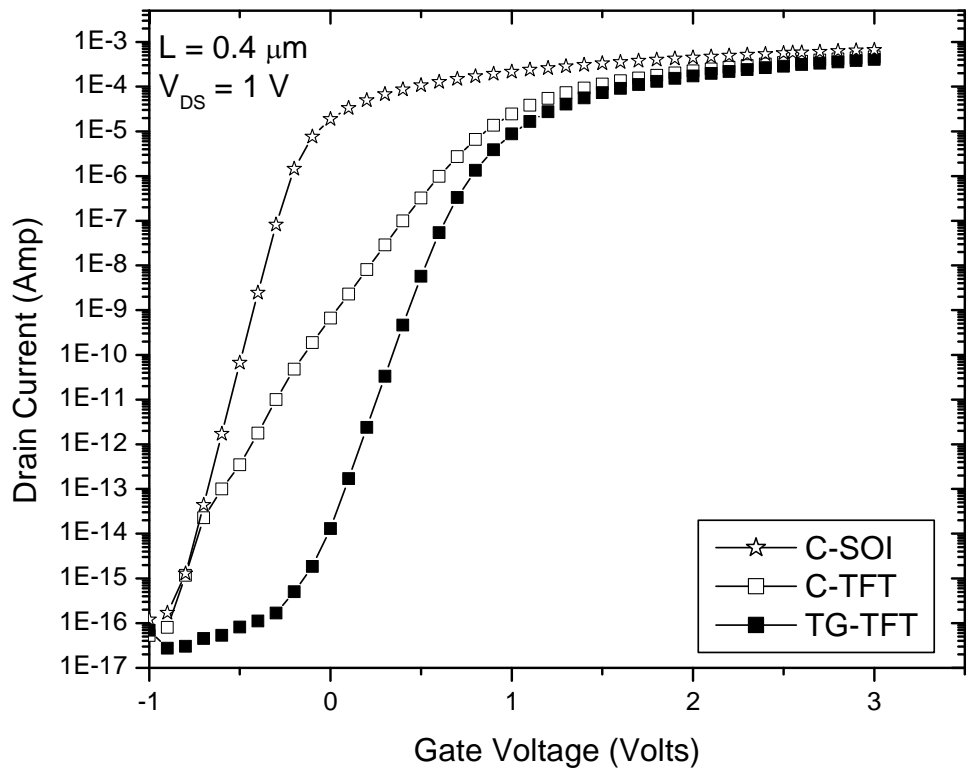

Fig. 18 (b)

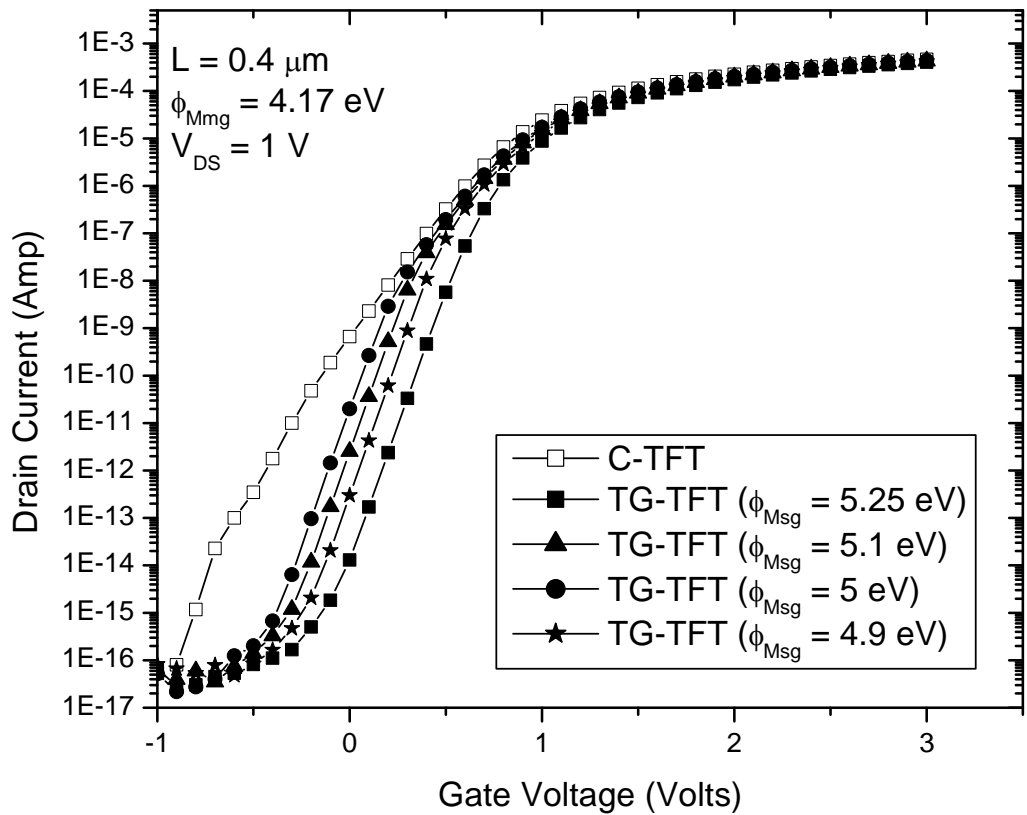

Fig. 19

# Author Biographies

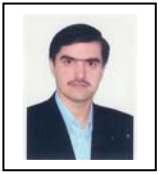

**Ali A. Orouji** (M'05) was born in Neyshabour, Iran in 1966. He received the B.S. and M.S. degrees in Electronic Engineering from the Iran University of Science and Technology (IUST) Tehran, Iran. He is currently working toward the Ph.D. degree in the Department of Electrical Engineering, Indian Institute of Technology, Delhi, India.

Since 1992, he has been working at the Semnan University, Semnan, Iran as a faculty member. His research interests are in modeling of SOI MOSFETs, novel device structures, and analog integrated circuits design.

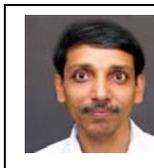

**M. Jagadesh Kumar** (SM'99) was born in Mamidala, Nalgonda District, Andhra Pradesh, India. He received the M.S. and Ph.D. degrees in electrical engineering from the Indian Institute of Technology, Madras, India. From 1991 to 1994, he performed post-doctoral research in modeling and processing of high-speed bipolar transistors with the Department of Electrical and Computer Engineering, University of Waterloo, Waterloo, ON, Canada. While with the University of Waterloo, he also did research on amorphous silicon TFTs. From July 1994 to December 1995, he was initially with the Department of Electronics and Electrical Communication Engineering, Indian Institute of Technology, Kharagpur, India, and then joined the Department of Electrical Engineering, Indian Institute of Technology, Delhi, India, where he became an Associate Professor in July 1997 and a Full Professor in January 2005. His research interests are in VLSI device modeling and simulation for nanoscale applications, integrated-circuit technology, and power semiconductor devices. He has published more than 105 articles in refereed journals and conferences in the above areas. His teaching has often been rated as outstanding by the Faculty Appraisal Committee, IIT, Delhi.

Dr. Kumar is a Fellow of Institution of Electronics and Telecommunication Engineers (IETE), India. He is on the editorial board of *Journal of Nanoscience and Nanotechnology*. He has often reviewed for different journals including *IEEE Trans. on Electron Devices, IEEE Trans. on Device and Materials Reliability, IEE Proc. on Circuits, Devices and Systems, Electronics Letters and Solid-state Electronics.* He was Chairman, Fellowship Committee, *The Sixteenth International Conference on VLSI Design,* January 4-8, 2003, New Delhi, India. He was Chairman of the Technical Committee for High Frequency Devices, *International Workshop on the Physics of Semiconductor Devices,* December 13-17, 2005, New Delhi, India. He is a member of the International Program Committee of *2006 International Conference on Computing in Nanotechnology (CNAN'06)*, June 26-29, 2006, Las Vegas, USA.